\documentclass[twocolumn]{aastex63}
\usepackage{float}
\usepackage[utf8]{inputenc}

\usepackage{graphicx}
\usepackage{hyperref}
\usepackage{amsmath}
\usepackage{color}
\usepackage{xcolor}
\usepackage{enumitem}

\hypersetup{colorlinks=true, citecolor=cyan, urlcolor = blue, linkcolor = blue}

\newcommand{\rev}[1]{\color{black}{#1}\color{black}}
\newcommand{\needrev}[1]{\color{green}{#1}\color{black}}

\shorttitle{Quasars with Periodic Variability}
\shortauthors{Witt et al.}

\received{October 18, 2021}
\revised{June 29, 2022}
\accepted{July 20, 2022}
\submitjournal{\apj}

\begin{document}
\title{
Quasars with Periodic Variability: 
Capabilities and Limitations of Bayesian Searches for Supermassive Black Hole Binaries in Time-Domain Surveys}

\correspondingauthor{Caitlin A. Witt}
\email{caw0057@mix.wvu.edu}

\author[0000-0002-6020-9274]{Caitlin A. Witt}
\affiliation{Department of Physics and Astronomy, West Virginia University, P.O. Box 6315, Morgantown, WV 26506, USA}
\affiliation{Center for Gravitational Waves and Cosmology, West Virginia University, Chestnut Ridge Research Building, Morgantown, WV 26505, USA}

\author[0000-0003-3579-2522]{Maria Charisi}
\affiliation{Department of Physics and Astronomy, Vanderbilt University, 2301 Vanderbilt Place, Nashville, TN 37235, USA}

\author[0000-0003-0264-1453]{Stephen R. Taylor}
\affiliation{Department of Physics and Astronomy, Vanderbilt University, 2301 Vanderbilt Place, Nashville, TN 37235, USA}

\author[0000-0003-4052-7838]{Sarah Burke-Spolaor}
\affiliation{Department of Physics and Astronomy, West Virginia University, P.O. Box 6315, Morgantown, WV 26506, USA}
\affiliation{Center for Gravitational Waves and Cosmology, West Virginia University, Chestnut Ridge Research Building, Morgantown, WV 26505, USA}
\affiliation{Canadian Institute for Advanced Research, CIFAR Azrieli Global Scholar, MaRS Centre West Tower, 661 University Ave. Suite 505, Toronto ON M5G 1M1, Canada}

\begin{abstract}

Supermassive black hole binaries (SMBHBs) are an inevitable consequence of galaxy mergers. At subparsec separations, they are practically impossible to resolve, and the most promising technique is to search for quasars with periodic variability.
However, searches for quasar periodicity in time-domain data are challenging due to the stochastic variability of quasars.
In this paper, we used Bayesian methods to disentangle periodic SMBHB signals from intrinsic damped random walk (DRW) variability in active galactic nuclei light curves. We simulated a wide variety of realistic DRW and DRW+sine light curves. Their observed properties
are modeled after the Catalina Real-time Transient Survey (CRTS) and expected properties of the upcoming Legacy Survey of Space and Time (LSST) from the Vera C. Rubin Observatory.
Through a careful analysis of parameter estimation and Bayesian model selection, we investigated the range of parameter space for which binary systems can be detected. We also examined which DRW signals can mimic periodicity and be falsely classified as binary candidates. We found that periodic signals are more easily detectable if the period is short or the amplitude of the signal is large compared to the contribution of the DRW noise. We saw similar detection rates both in the CRTS and LSST-like simulations, while 
the false-detection rate depends on the quality of the data and is minimal in LSST. 
Our idealized simulations provide an excellent way to uncover the intrinsic limitations in quasar periodicity searches and set the stage for future searches for SMBHBs. 
\end{abstract}

\section{Introduction}

Supermassive black hole binaries (SMBHBs) should form frequently in the aftermath of galaxy mergers \citep{Haehnelt2002}. However, the evolution from this initial stage to the formation of a bound binary and the final coalescence is complex. After the galaxy merger, the supermassive black holes (SMBHs) hosted in the cores of their parent galaxies sink to the center of the created galactic remnant through dynamical friction. At scales of a few parsecs, stellar scatterings and interactions with ambient gas continue shrinking the binary orbit. If these processes remove sufficient energy and angular momentum so that the binary efficiently overcomes the ``final-parsec problem," then gravitational waves (GWs) dominate the binary decay and drive the binary to the final merger \citep{Begelman1980, Colpi2014, DeRosa2019}.

The most massive binaries (total mass of $10^8-10^{10}M_{\odot}$) emit GWs at low frequencies (few to hundreds of nanohertz). GWs in this frequency band can be detected by pulsar timing arrays (PTAs),  
and offer one of the only direct probes to SMBHBs at close (roughly milliparsec) separations \citep{Burke-Spolaor2019,2019BAAS...51c.336T}. Electromagnetic observations can also infer the existence of a SMBHB, 
and provide a unique probe of the binary's environment \citep{2021arXiv210903262B}. 
The detection of GWs along with  associated electromagnetic counterparts will mark the beginning of multimessenger astrophysics in the low-frequency regime \citep{2019BAAS...51c.490K}. 
%
In fact, multimessenger techniques are already being developed. Incorporating information from SMBHB candidates in GW searches allows us to place tighter constraints on the SMBHB chirp mass \citep{3c66b}, and can boost the detectability of the candidate in a typical ``blind" search \citep{2021arXiv210508087L}. 

Closely separated SMBHBs in the GW regime may be detected as active galactic nuclei (AGNs) or quasars with periodic variability \citep{2009ApJ...700.1952H}. Previous studies have demonstrated a link between AGNs and galaxy mergers; this follows naturally from the idea that the mergers bring significant amounts of gas to the central regions of the post-merger galaxies, which may actively accrete onto the SMBHs, triggering AGN activity \citep{2018PASJ...70S..37G}. Similarly, binaries are expected to be surrounded by significant amounts of gas, which can give rise to bright quasar-like electromagnetic emission  \citep{2002ApJ...567L...9A,2012MNRAS.420..705T,2021arXiv210903262B}.  

Specific predictions for periodic variability in binary AGNs has been demonstrated in multiple hydrodynamical simulations of binaries embedded in gaseous disks \citep{2008ApJ...672...83M,2012A&A...545A.127R,2013MNRAS.436.2997D,2014ApJ...783..134F}. The consensus of these simulations is that the binary carves out a central cavity, i.e. a region of low-density gas. As the binary orbit perturbs the edge of this cavity (especially the secondary SMBH, which moves closer to the edge), it pulls streams of gas inwards. Periodic accretion onto the SMBHs 
from these streams 
may produce periodic brightness fluctuations. Another mechanism that produces periodic variability is relativistic Doppler boosting \citep{Dorazio2015Nature,Tang+2018}. 
Some of the gas that penetrates the cavity ends up bound to the SMBHs, forming mini-disks which orbit with relativistic speeds. The emission from these mini-disks may be periodically boosted (and dimmed), even if the rest-frame luminosity is constant. This signature is prominent for unequal-mass binaries orbiting close to edge-on, where the emission of the faster moving secondary ---which is also typically brighter---dominates the variability.

In recent years, vast photometric databases of time-domain surveys have provided light curves for large samples of AGNs, which are ideal for searches of SMBHBs. Numerous candidates have been identified from systematic searches in
optical surveys, such as the Catalina Real-time Transient Survey \citep{Graham2015}, the Palomar Transient Factory \citep{Charisi2016}, the  Panoramic Survey Telescope and Rapid Response System \citep{2019ApJ...884...36L}, and the Dark Energy Survey \citealt{2020MNRAS.499.2245C}.
However, AGNs also have intrinsic stochastic variability, which makes periodicity identification quite difficult. AGN variability is successfully modeled by a ``damped random walk'' (DRW) model, which takes the form of a red-noise process at high frequencies but a white-noise process at low frequencies \citep{kozlowski2010, macleod}. This intrinsic noise is impressively capable at mimicking periodicity, particularly in sparsely sampled or short-baseline time series \citep{vaughn}. So far, studies have focused on additional signatures for the binary nature of candidates, such as multiwavelength Doppler boost \citep{Dorazio2015Nature,2018MNRAS.476.4617C,2020MNRAS.496.1683X}, periodicity with multiple components \citep{Charisi2015}, X-ray spectral excess \citep{2020ApJ...900..148S} and distorted radio jets \citep{Kun2015,Mohan2016}.

However, multiwavelength follow-up monitoring of candidates is demanding and such studies will be impractical (if not impossible) in the upcoming generation of surveys like the Legacy Survey of Space and Time (LSST) of the Vera C. Rubin Observatory \citep{2009arXiv0912.0201L}. LSST is expected to observe over 20
million of quasars, delivering an unprecedented data set for quasar periodicity searches both in terms of quality and quantity. If we extrapolate the detection rate of SMBHB candidates in the current time-domain surveys ($\sim 1/1000$) to LSST, we expect several thousands of candidates. However, we know that these samples likely contain many false detections (as demonstrated by their tension with the GW background limits when extrapolated to a full binary population; \citealt{Sesana2018}). On the other hand, theoretical models predict that hundreds of genuine binaries should be detectable in LSST \citep{2019MNRAS.485.1579K,2021MNRAS.506.2408X,2021arXiv210707522K}.
Because of this, the time is ripe to develop a careful model selection in order to reliably identify binary candidates.

In this work, we explore the capabilities and limitations in identifying quasars with periodic variability in the data sets of the upcoming decade. 
We simulate idealized AGN light curves that contain DRW noise with realistic parameters, while a subset of those contains sinusoidal variations on top of the DRW noise. We construct a pipeline that employs Bayesian model-selection and parameter-estimation to identify periodic signals (i.e. the binary candidates) in our sample, and constrain their parameters. Finally, we quantify our ability to select genuine binaries and the degree of contamination with false detections.

This paper is laid out as follows. In \autoref{sec:methods}, we describe the 
methodology for creating simulated light curves, as well as the Bayesian parameter estimation and model selection methods. In \autoref{sec:results}, we  examine the efficacy of our Bayesian pipeline, as well as present a statistical analysis of this efficacy across the simulated SMBHB population. In \autoref{sec:conclusions} we present the conclusions we can draw from our analysis. These include the following key findings:
\begin{itemize}[noitemsep, nolistsep]
    \item Our method can recover orbital periods extremely accurately (even very long or short values), provided the signal is of sufficient strength. The detectability of periodicity also depends on the amplitude of the sinusoid and the contribution of the DRW noise.
    \item While a DRW process can mask some sinusoids in current surveys, the false-positive rate is very low for LSST, and thus it is expected to deliver reliable candidates.
    \item Particular combinations of DRW and sinusoidal parameters are more likely to mask a signal than others. This will help inform future analyses as we attempt to confront the massive data volume that will be produced by LSST.
\end{itemize}
Finally, in \autoref{sec:discussion}, we discuss caveats of our method, future improvements, and the prospects of multimessenger observations of binaries.
This work presents a necessary first step in preparation for the flood of SMBHB candidates in the upcoming Rubin era.

\section{Methods} \label{sec:methods}

\begin{figure*}
    \centering
    \includegraphics[width =1\textwidth]{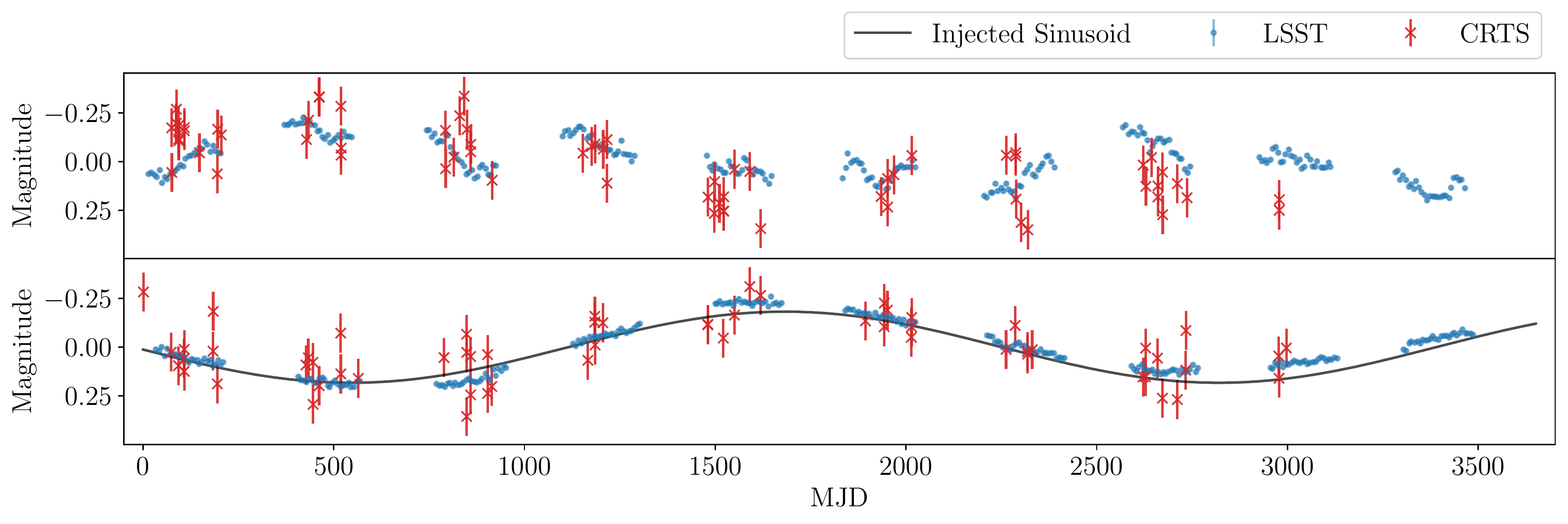}
    \caption{An example of simulated light curves containing a DRW process (top panel) and a DRW process plus a sinusoid (bottom panel). The sinusoid is shown in the solid black curve, while the simulated data for a CRTS-like and LSST-like survey are shown as red X's and blue points, respectively. Depending on the DRW and sinusoid parameters, it is possible for these two models to produce deceptively similar results. }
    \label{fig:lightcurve}
\end{figure*}

As mentioned above, identifying periodicity in quasars is challenging because of the intrinsic stochastic variability of quasars, the relatively short observation baselines compared to the potential binary periods, and the noisy, irregular data. Our goal is to explore the variety of binary signals (e.g., range of periods, amplitudes) that can be detected in current and upcoming time-domain surveys. We also aim to assess the expected false-positive rate in systematic searches for quasar periodicity. For this, we simulate typical quasar light curves with realistic DRW noise properties as well as SMBHB light curves which include sinusoidal signals with a variety of periods and amplitudes on top of DRW noise. We chose to model the binary signal with a pure sinusoid both for simplicity and because previous searches for quasars with periodic variability have focused on quasi-sinusoidal signals. We construct a periodicity-detection pipeline that employs a Bayesian model parameter estimation and selection between a DRW and DRW+sine model, and apply it in a wide range of simulated light curves. Below we describe the light-curve simulations and the periodicity-detection method.

\subsection{Simulated Data}\label{subsec:sims}

To ensure that our analysis was realistic, we constructed our simulated light curves with properties that reflect the observational capabilities of ongoing and planned time-domain surveys. In particular, we assessed the detectability of SMBHBs in current surveys, using CRTS-like light curves, whereas for future surveys we used idealized LSST-like light curves. Each survey has a distinct observing strategy (depending on their primary scientific objectives), which defines the average cadence (frequency of observations) and observation baseline (length of light curve). Additionally, each survey has a limiting depth, which depends on the size of the telescope used and the exposure time of the typical observation. This defines the photometric uncertainty, which is typically a function of apparent magnitude; dimmer sources have larger photometric errors and vice versa. \rev{For CRTS, we used observed light curves as the basis for our simulations, whereas for LSST we simulated light curves based on expectations for the cadence, as described below. } 
However, a future study should address this and other limitations, as discussed in \autoref{sec:discussion}. 

In order to construct the CRTS-like light curves, we extracted 10,000 AGN light curves spread across the sky from the online database.\footnote{\url{http://nesssi.cacr.caltech.edu/DataRelease/}} We examined the sampling pattern in this set, which turns out to be similar for most light curves. In particular, for each night the source was observed, the light curves contain clusters of four successive data points, then the next set of observations is taken about one week to one month later for as long as the source is observable (for about six months). Subsequently, there is a significant gap of no data for about six months, e.g., when the source is obstructed by the sun or below the horizon of the specific telescope, and then the pattern roughly repeats. \rev{For the sampling of the simulated light curves, we selected a random subset from that sample and used them as templates for our simulations}. These light curves have an average span of $\sim$20 days between successive nights of observations, gaps of $\sim$200 days, and a total observation baseline of \rev{between 7 -- 11 }yr. This gives an effective cadence (observation baseline divided by the number of data points) of 46 days. Since very short term variations are not relevant for our study, this calculation did not include multiple observations within the same night. 
\rev{The simulated light curves have a range of mean magnitudes and mean photometric uncertainties. We also calculated the mean magnitude and mean magnitude error for the ensemble of simulations, for comparison with LSST below, which we found to be, on average, $\sim$18 and 0.1, respectively.}


For LSST, the nominal duration is set to 10 yr, but the observing strategy is not finalized yet \citep{LSSTObservingStrategyWhitePaper}. The majority of time will be spent on the deep-wide-fast survey mode, which will cover a 18,000 deg$^2$ footprint with a regular cadence. We set our simulation cadence at a conservative value of seven days, while actual observations may repeat every five or even three nights. We note, however, that LSST will rotate between six filters, and successive observations will provide data in different photometric bands. We do not take this into account in our simulations, but we discuss this caveat further in \autoref{sec:discussion}. Since the observations will not repeat in exactly seven day increments, we create a linear grid of time stamps separated by seven days and add 
Gaussian noise with a standard deviation of one day. 
\rev{We then introduce gaps with a duration of approximately six months. Even though gaps are inevitable for ground-based observations, our choice here is rather conservative, since it is highly likely that LSST will have longer epochs compared to previous time-domain surveys, with eight months of uninterrupted observations followed by gaps of four months. }
\rev{Finally, we distribute the average photometric error about a value of 0.01 mag, which corresponds to a mean $r$-band magnitude of $\sim$21, as we expect a sizeable sample of AGNs with magnitudes of 21 and higher } \citep{lsst_science_drivers}.
See \autoref{tab:surveys} for a summary of the parameters of the simulated light curves for each survey.

\begin{table}
    \centering
    \begin{tabular}{ccccc}
        \hline
         {Survey} & Baseline & Cadence  & Mean  & Mean Phot.  \\
                  & (years) &  (days) & Magnitude & Error (mag)  \\
         \hline
         \hline
         CRTS & 7 -- 11 & 46 & 18 & 0.1 \\
         LSST & 10& 7 & 21 & 0.01 \\
         \hline
    \end{tabular}
    \caption{Average parameters for each survey used to create simulated data sets.}
    \label{tab:surveys}
\end{table}

With the observed properties of the time series, we proceeded to simulate DRW and DRW+sine light curves following the steps from \citet{Charisi2016}. The power spectral density (PSD) function of DRW is 
\begin{equation}\label{eq:psd}
    P(f)=\frac{4 \sigma^{2} \tau}{1+(2 \pi \tau f)^{2}},
\end{equation}
where $\sigma^2$ is the variance of the light-curve data points, $\tau$ is a characteristic DRW timescale, and $f$ is the Fourier-space frequency. With the inverse Fourier transform of the PSD, we generate evenly sampled light curves (with $\Delta t=1\,day$) using the prescription from \citet{1995A&A...300..707T}, included in the python package \texttt{astroML} \citep{astroML,astroMLText}. We downsample the data to match the desired sampling pattern of the survey setup described above and in \autoref{tab:surveys}. Next, we add Gaussian errors with zero mean and standard deviation equal to the average photometric uncertainty of the respective survey (\autoref{tab:surveys}).

For the set of simulations that also include SMBHB signals, we inject a sinusoid on top of the DRW light curve. This signal has the form
\begin{equation}
    \mathbf{s}(t) = A \sin\left(\frac{2 \pi}{P}(t-t_0)\right),
\end{equation}
where $A$ is the amplitude in magnitudes, $P$ is the period of the sinusoid, and $t_0$ is a reference time. Both the period and the amplitude of the sinusoid can be linked to the parameters of the binary; the observed period is typically the redshifted orbital period of the binary, and if the periodicity is produced by relativistic Doppler boost, the amplitude $A$ depends on the line-of-sight velocity of the secondary SMBH.
Example time series with a simulated DRW-only process, and DRW+sine, can be seen in \autoref{fig:lightcurve}, where with blue (and red) data points we show the LSST-like (and CRTS-like) light curves.

We generated DRW and DRW+sine light curves for a wide variety of these five input parameters ($P,~A,~t_0,~\sigma$, and $\tau$). First, in order to test the Bayesian pipeline's ability to recover the model parameters, we choose values across an extreme range of $\tau$ corresponding to those used in \citet{kozlowski2017}.
Injected values are randomly selected from the range $\tau = [10^{-3}T, 15T]$, where $T$=10\,yr is the nominal LSST observation baseline. 
\rev{This wide range of $\tau$ values is intended to ensure that our model-selection method will be capable of analyzing data across a variety of future surveys, which may expand our knowledge of the inherent AGN distribution. }
However, 
for the model-selection analysis we restrict the values of $\tau$ to a realistic distribution derived from those presented in \citet{macleod}.
For $\sigma$, we draw values from a log-uniform distribution ranging from $[-1.6,-0.25]$, corresponding to a range of greater than an order of magnitude in $\sigma$, to encompass a wide range of DRW variability amplitudes similar to the range presented in \citet{macleod}. \rev{We note that \citet{macleod} reported a weak correlation between $\sigma$ and $\tau$, recently updated by \citet{2021ApJ...907...96S}. Our simulations randomly drawing from uniform priors do not incorporate this correlation, which may contain biases due to the limited sample size \citep{graham_2017}. However, we plan to address this limitation in a future study.}

The periods of the injected sinusoids range from 30 days to 10 yr. The maximum value is set by the LSST baseline, so that at least one full orbital cycle is observed. This wide range of periods covers all the potential SMBHBs that have GW frequencies detectable by PTAs. However, it does not include very high frequency SMBHBs possibly detectable by the Laser Interferometer Space Antenna (LISA; \citealt{2021MNRAS.506.2408X}), which are expected to have periods of only a few days ($P<1-2$\,days). In \autoref{sec:discussion}, we explore whether such short-period binaries need a distinct strategy for detection, such as accounting for filter alternation and combining the multiband data in a single light curve.
Previous studies have required that at least 1.5 cycles (or more) of the periodicity be observed within the available baselines. We relax this requirement to assess the ability to recover binaries in this regime and the resulting contamination with false positives. This is significant, since binaries evolve slower at large separations, and long-period binaries are expected to be more common.
The reference time is set to any time between 0 and the \rev{maximum allowed sinusoid period, which can also be described by a phase in the range [0, 2$\pi$]. }
The amplitude is set to a value in the range 
$[0.05, 0.5]$ mag.
These distributions of simulated values are summarized in \autoref{tab:priors}.

\subsection{Likelihood and Sampling Methods}\label{ss:likelihood_mcmc}

For the DRW process defined in \autoref{eq:psd}, the covariance matrix $S$ that determines the correlation between two data points at times $t_i$ and $t_j$ is given by 
\begin{equation}\label{eq:cov}
    S_{i j}=\sigma^{2} \exp \left(-\frac{\left|t_{i}-t_{j}\right|}{\tau}\right),
\end{equation}
where $\sigma^2$ and $\tau$ are the same values defined above. The full covariance matrix is $C = S+N$, where $N = \mathrm{diag}(\sigma_{\mathrm{err}}^2)$ is the noise covariance matrix with $\sigma_{\mathrm{err}}$ the survey's photometric error. The DRW likelihood function marginalized over the mean of the light curve is given by
\begin{equation}\label{eq:likelihood}
\begin{split}
    P(\mathbf{y} \mid \mathbf{p}) \propto  |C|^{-1 / 2}  & \left|L^{T} C^{-1} L\right|^{-1 / 2} \\ & \times \exp \left(-\frac{\mathbf{y}^{T} C_{\perp}^{-1} \mathbf{y}}{2}\right),
\end{split}
\end{equation}
with $\mathbf{y}$ the vector of the data (observed magnitudes) and $L$ a vector of ones with a length equal to the number of data points, and \begin{equation}
    C_{\perp}^{-1}=C^{-1}-C^{-1} L\left(L^{T} C^{-1} L\right)^{-1} L^{T} C^{-1}.
\end{equation} 
For a detailed derivation we refer the reader to \citet{kozlowski2010}.
The likelihood function for the DRW+sine model is given by
\begin{equation}\label{eq:likelihood_sine}
\begin{split}
    P(\mathbf{y} \mid \mathbf{p}) \propto  &|C|^{-1 / 2}   \left|L^{T} C^{-1} L\right|^{-1 / 2} \\ & \times \exp \left(-\frac{(\mathbf{y}-\mathbf{s})^{T} C_{\perp}^{-1} (\mathbf{y}-\mathbf{s})}{2}\right),
\end{split}
\end{equation}
with 
$s$ a vector of the sinusoid $s(t) = A~ \mathrm{sin}~(2\pi/P(t-t_0))$ sampled at the observed times.

We utilize \rev{Bayesian } methods for both parameter estimation and model selection. \rev{In particular, we sample the likelihood using a nested-sampling Monte Carlo algorithm, MLFriends \citep{mlfriends1, mlfriends2}, using the \texttt{Ultranest}\footnote{\url{https://johannesbuchner.github.io/UltraNest/}} package \citep{ultranest}. This nested sampler efficiently explores the entire parameter space and avoids effects induced by the unevenly sampled time series. }

In general, we use relatively uninformative priors for our \rev{Bayesian } analyses
(either uniform or log-uniform), as summarized in \autoref{tab:priors}.
The priors typically 
\rev{span the entire range of the }
distributions of simulated parameters described in \autoref{subsec:sims}. 
We chose flat priors to avoid introducing potential biases and to emulate an uninformed systematic search. More informative priors could be imposed for the DRW parameters; for example,  \citet{macleod} found that $\sigma$ and $\tau$ are correlated with the properties of the AGN (e.g., the SMBHB mass, the luminosity, etc). Since in our simulated light curves we did not vary luminosity-related parameters (e.g., the observed magnitude) a fairly unrestricted search is more appropriate.

\begin{table*}
    \centering
    \begin{tabular}{ccc}
        \hline
        {Parameter} & Simulation Distribution & Prior \\
        \hline
        \hline

         $\mathrm{log}_{10}\sigma$ & Log-Uniform[-1.6, -0.25]  & Log-Uniform[-1.6, -0.25]\\
         $\mathrm{log}_{10}\tau $ (Wide Range)& Log-Uniform[0.56, 4.73] & Log-Uniform[0.56, 4.73] \\
         $\mathrm{log}_{10}\tau $ (Realistic Distribution) & SkewNorm(3.0, 0.5, -1.4) & Log-Uniform[0.56, 4.73] \\
         $\mathrm{log}_{10}P $ & Log-Uniform[1.5, 3.5] & Log-Uniform[1.5, 3.5] \\
         $A$ & Uniform[0.05, 0.5] & Uniform[0.05, 0.5] \\
         $t_0$ & Uniform[0, 3650] & Uniform[0.05, 0.5] \\
         \hline
    \end{tabular}
    \caption{Simulation ranges for each of our five parameters, and prior shapes and ranges for our \rev{analysis}. Note that for model selection analyses, we simulate the realistic distribution of $\tau$ values derived from \citet{macleod}.}
    \label{tab:priors}
\end{table*}

For each simulated light curve, we performed the analysis for two models (DRW and DRW+sine): the first uses the DRW likelihood from \autoref{eq:likelihood} to search over only the two DRW parameters, $\sigma$ and $\tau$, and the second uses \autoref{eq:likelihood_sine}, which also searches over the sinusoid parameters. 
From \rev{the resulting posterior distributions}, we estimated the values of the two (or five) parameters that are most likely given each simulated light curve. The posterior distributions provided both median values and uncertainties for the parameters. From the posteriors we also determined the value for set of parameters that maximized the likelihood. For each simulation, we also calculated a signal-to-noise ratio (S/N), where
\begin{equation}
    \mathrm{S/N}^2 = \mathbf{s}^T \cdot C^{-1} \cdot \mathbf{s}.
\end{equation}
Here, $\mathbf{s}$ is the vector containing the input signal and sampled at the simulated time stamps, and $C^{-1}$ is the inverse of the DRW covariance matrix (\autoref{eq:cov}). 

We used the outcome of the two searches (DRW and DRW+sine) to perform Bayesian model selection using the Bayes information criterion (BIC)
\begin{equation}\label{eq:bic}
    \mathrm{BIC}=k \ln (n)-2 \ln (\widehat{L}),
\end{equation}
where $k$ is equal to the number of free parameters, $n$ is equal to the number of data points in the light curve, and $\widehat{L}$ is the maximum likelihood value \citep{liddle}. The BIC provides a simple metric through which to compare our two models, and avoids overfitting the data by accounting for the number of parameters in the model. When selecting among multiple models, the one with the smallest BIC is usually preferred.
Here we selected the preferred model by comparing the BIC values for the DRW-only search to that of the DRW+sine search by introducing
\begin{equation}
\Delta \mathrm{BIC} = \mathrm{BIC_{DRW}} -  \mathrm{BIC_{DRW+sine}}.
\end{equation}
A lower value of $\Delta \mathrm{BIC}$ indicates more support for the DRW+sine model. In general, evidence for the DRW+sine model can be considered positive for $-2~>~ \Delta \mathrm{BIC}~>~-6$, and strong for $\Delta \mathrm{BIC}~<~-6$ \citep{kass}. \rev{The $\Delta \mathrm{BIC}$ can also be used to estimate a Bayes factor, $\mathcal{B}_{10}$ for the model comparison, where 
\begin{equation}
    \mathcal{B}_{10} = e^{ \left ( - \frac{1}{2} \Delta \mathrm{BIC} \right)}.
\end{equation} } 
Here, we defined our threshold to claim a detection of a sinuosidal signal as $\Delta \mathrm{BIC} = -2$. Using this threshold, we sorted each result into one of four categories:
\begin{itemize}[noitemsep, nolistsep]
    \item \textbf{True positive}: A sinusoid was injected and the DRW+sine model was preferred.
    \item \textbf{False negative}: A sinusoid was injected, yet the DRW-only model was preferred.
    \item \textbf{False positive}: No sinusoid was injected, yet the DRW+sine model was preferred
    \item \textbf{True negative}: No sinusoid was injected, and the DRW-only model was preferred.
\end{itemize}

In an idealized search we would have only true positives/negatives and no false positives/negatives, but typically one needs to compromise and balance the rate of detection of true signals with the contamination of a few false positives.
One of the main goals of this analysis is to constrain these rates for current and future survey capabilities. We note, however, that these rates refer to our specific method of Bayesian model selection and cannot be extended to existing samples of SMBHB candidates, since these candidates were selected with completely different methods, as we explain in \autoref{sec:discussion}.

\section{Results} \label{sec:results}
We assessed our ability to identify periodicity in AGN light curves by simulating DRW and DRW+sine light curves and performing \rev{a } Bayesian model selection. First, we tested how \rev{our } algorithm performs in constraining the parameters of each model independently. Subsequently, we determine the performance of the model-selection method by calculating the true- and false-positive rates, and characterizing their dependence on the signal and noise parameters.

\subsection{Parameter estimation}\label{subsec:params}
\subsubsection{Damped Random Walk Model}
\begin{figure}
    \centering
    \includegraphics[width = 1\columnwidth]{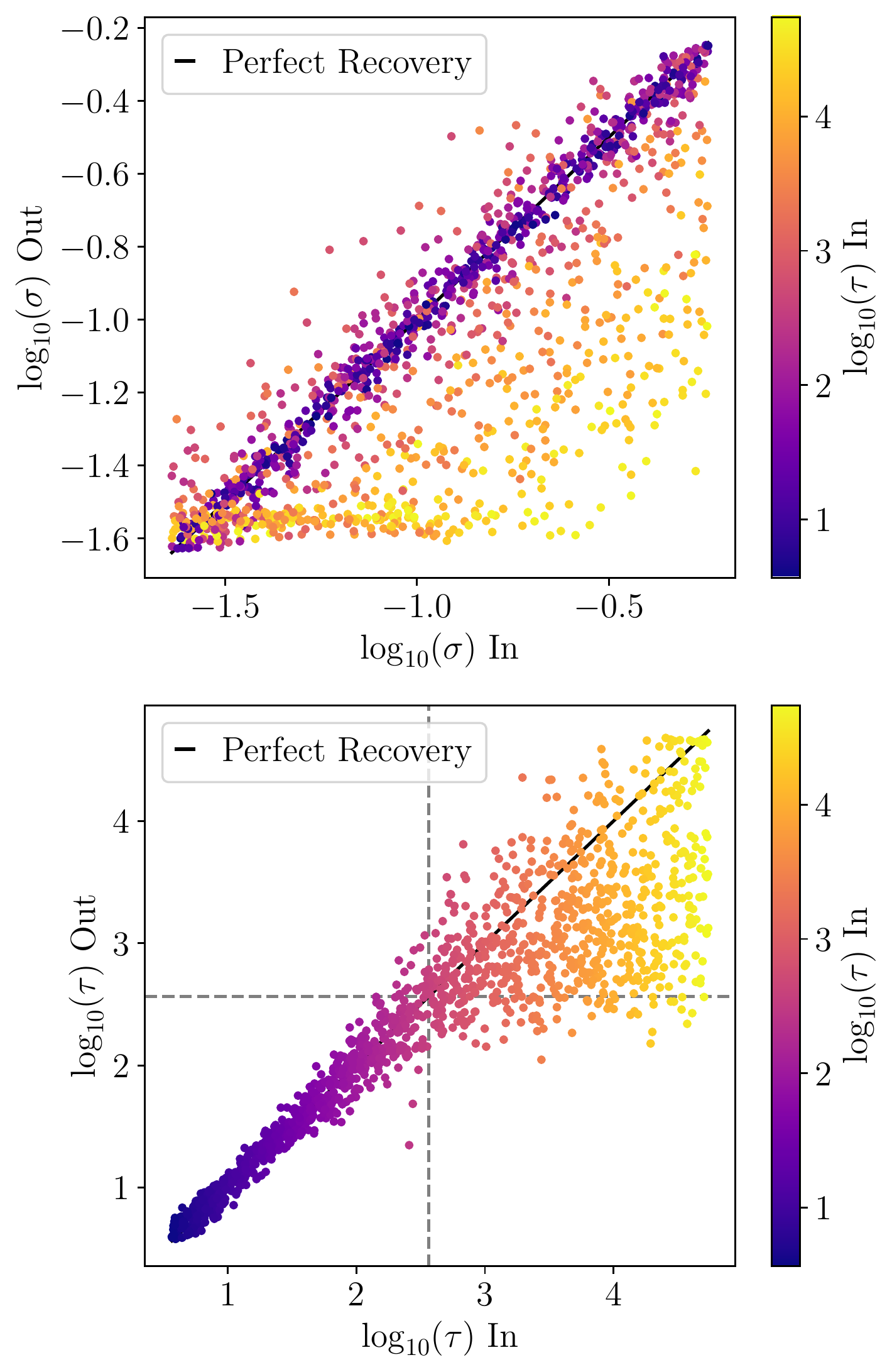}
    \caption{Parameter estimation capability of our methods for LSST-like light curves with a DRW process, colorized by the simulated value of $\tau$. For low values of $\tau$, both parameters are recoverable as expected. However, for very high values of $\tau$, both are unlikely to be constrained accurately.}
    \label{fig:drw_inout}
\end{figure}

For both our CRTS-like and LSST-like simulations, we simulated \rev{1500 } DRW light curves with properties as described in \autoref{subsec:sims} and conducted the nested-sampling analysis for the DRW likelihood (\autoref{eq:likelihood}). In \autoref{fig:drw_inout} we show the median values of $\sigma$ and $\tau$ as a function of the respective input values for the LSST light curves. We note that the parameter estimation for the DRW model shows very similar trends for the CRTS-like light curves.

We saw that low values of $\tau$ are recovered accurately, while high values were poorly constrained. This is a known limitation in DRW studies. For instance, \citet{kozlowski2017} found that for $\tau$ to be well recovered, the baseline of the light curve must be at least ten times greater than $\tau$ ($\tau\leq 10T$). 
In that study, the authors demonstrated this effect with simulated light curves for a fixed parameter $\sigma$. Here, by varying the values of $\sigma$ for each simulation, we demonstrate that this limitation affects the recovery of $\sigma$ as well. In \autoref{fig:drw_inout} we colorized the data points by the input value of $\tau$. We observed that for light curves with large values of $\tau$, where $\tau$ is not constrained (orange-yellow points), the algorithm fails to recover the input value of $\sigma$. On the other hand, for light curves with small values of $\tau$ (purple points), the recovery of both $\sigma$ and $\tau$ is very accurate.

\subsubsection{DRW+Sine Model}

Once we confirmed that the DRW parameters can be recovered by \rev{our } methodology (within already known limitations), we expanded our search to also include the sinusoidal signal representing an SMBHB. We repeated the \rev{1500 } simulations of both CRTS- and LSST-like DRW light curves, and added a randomly generated sinusoid to the data. This was then searched with a five-parameter \rev{model } using the DRW+sine likelihood from \autoref{eq:likelihood_sine}. 

\begin{figure*}
    \centering
    \includegraphics[width = 1\textwidth]{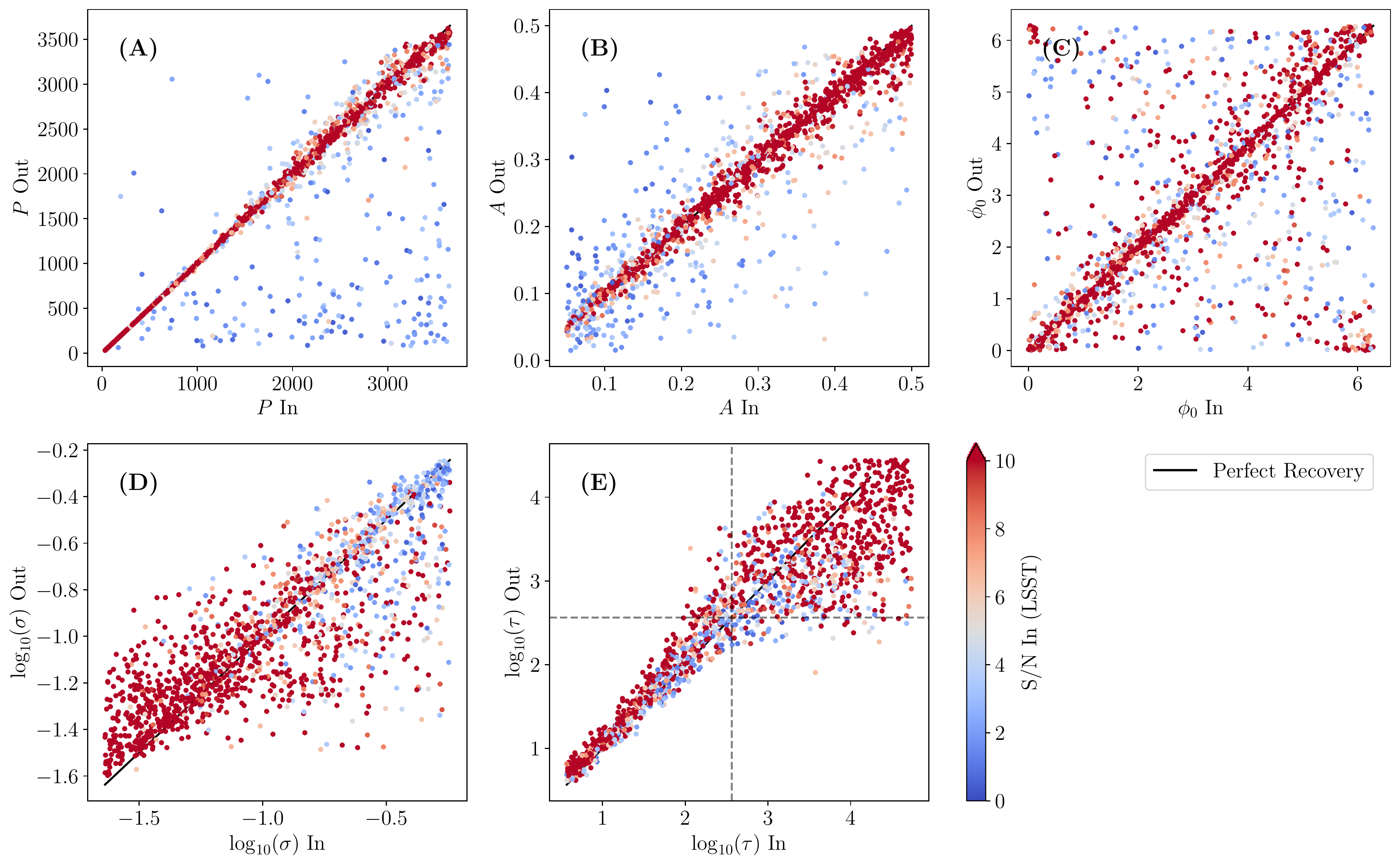}
    \caption{Recovery capabilities for the three sinusoid parameters ($P$, $A$, and $\phi_0$) and two DRW parameters ($\sigma$, and $\tau$), as demonstrated by our LSST-like light-curve simulations. The color axis represents the S/N of the input signal, with red points being strong signals (S/N$>$5). The sinusoids in these simulations were able to be recovered extremely accurately, and the DRW parameters were recovered to the extent we expect from noise-only simulations.}
    \label{fig:in_out_sine}
\end{figure*}

\autoref{fig:in_out_sine} summarizes the recovery capability of the DRW+sine model in LSST-like simulations, color-coded by the S/N of the input signal. We note that even though it is preferable to sample the likelihood in terms of a reference time $t_0$, we present results converted to an initial phase $\phi_0=2\pi t_0/P$ to avoid potential biases or correlations with the period. \rev{In general, for LSST-like simulations, we recovered sinusoids with S/N$>$5 extremely accurately. In particular, for 96\% of these signals, we recovered the injected periods and amplitudes, i.e., the injected value was within the 90\% credible region (5$^{\rm th}$ to 95$^{\rm th}$ percentile of the posterior distribution), } \rev{84\% } of these signals had all parameters recovered, whereas the DRW parameters were recovered for \rev{87\% } of signals.
\rev{It is important to note that for $\tau$, as discussed above, large input values are often underestimated. This is a likely cause for the lower recovery rate, when we consider sets of parameters that include the DRW parameters. }
\rev{The recovery capabilities are quite similar for our CRTS-like simulations, but due to the lower photometric accuracy there are fewer simulations with S/N$>$5. In this setup, 93\% of signals with S/N$>$5 are recovered with the input values of both their amplitudes and periods recovered accurately (within the credible region), 74\% of these had all parameters recovered accurately, and 80\% had the DRW parameters recovered accurately. }
This method is successful at recovering sinusoids with a wide range of injected parameters. It is important to note that the algorithm accurately recovered periods from 30 days to 10 years, and it was not required for all light curves to cover more than two cycles of the sinusoid for their parameters to be recoverable, as may be expected based on analyses by \citet{vaughn}.
We further explore the longer-period regime in \autoref{subsec:model_selection}.

The DRW parameters $\sigma$ and $\tau$ were recovered with the same accuracy as in the DRW-only search, even in the presence of the sinusoid. We also saw the same limitations in recovering long $\tau$ and resulting limitations in recovering $\sigma$ for this subset of light curves. However, our inability to constrain the DRW parameters in certain light curves was not propagated to the recovery of the parameters of the periodic signal. Additionally, the highest $\sigma$ values are at near the maximum of the observed quasar population, and will be fairly rare in reality \citep{macleod}. 

\subsubsection{Covariance of Timescales}\label{subsec:timescales}

\begin{figure}
    \centering
    \includegraphics[width = 1\columnwidth]{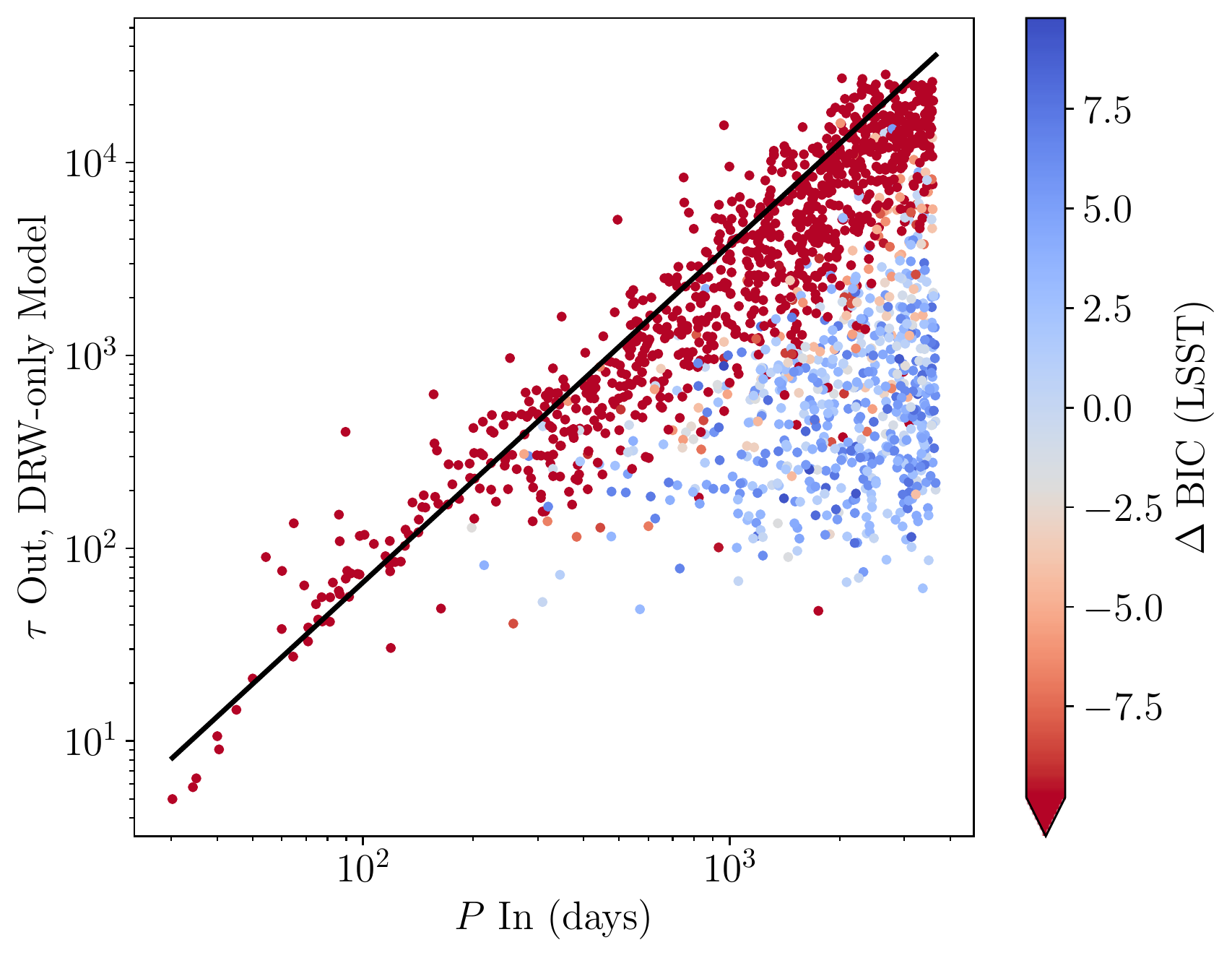}
    \caption{If a light curve is simulated to contain both a sinusoid and a DRW process, but is modeled with only noise, significant confusion can occur. The periods of recoverable sinusoids (red) can be confused for the DRW timescale $\tau$ if only noise is modeled.}
    \label{fig:tau_confuse}
\end{figure}

\rev{
We found that the stochastic DRW noise hinders the detection of the deterministic signal of a SMBHB. One potential reason is the covariance between the parameters of the signal and the noise. For instance, both the amplitude of the sinusoid and the DRW $\sigma$ determine the overall S/N of the light curves. Unsurprisingly, we saw that our ability to detect sinusoidal variability increases when $\sigma$ is small and $A$ is large, and vice versa. The covariance of the characteristic timescales $P$ and $\tau$ is less obvious, so we explore this issue in more detail below.

First, we examined our results for potential correlations when we fit for the incorrect model (i.e. injected DRW+sine using the DRW likelihood from \autoref{eq:likelihood}). Searching a light curve that has a sinusoid injected with a DRW-only model will result in a biased recovery of $\tau$, as can be seen in \autoref{fig:tau_confuse}, where the recovered $\tau$ value is related to the injected period. For LSST-like simulations, this was best fit with a linear function where }
\rev{\begin{equation}
    \log_{10}\tau = 1.75\left(\log_{10}P\right) -1.68. 
\end{equation} }
\rev{
However, for CRTS-like simulations, this covariance is best fit with a linear function:} 
\rev{\begin{equation}
    \log_{10}\tau=1.30\left(\log_{10}P\right) -1.25.
\end{equation}}
\rev{
\citet{kozlowski2010} found a similar effect when they applied the DRW formalism to periodic stellar light curves (e.g., their Fig. 12 and the related discussion). We also confirm their finding that these correlations are sensitive to the light-curve properties, since we find a different correlation in our CRTS-like and LSST-like light curves.
When $\tau$ is fit in conjuction with the periodicity (i.e in the DRW+sine model), this bias is resolved. 
}

\subsection{Model Selection}
\label{subsec:model_selection}
Next, we used a Bayesian model selection, described in detail in \autoref{sec:methods}, to select quasars with periodic variability. With simulated DRW and DRW+sine light curves, we traced the algorithm's effectiveness. We considered two distinct surveys (CRTS and LSST, reflecting current and future capabilities of time-domain surveys) to explore how the light-curve quality and properties affect the detection rates of this method. 

\begin{figure*}
    \centering
    \includegraphics[width = 1\textwidth]{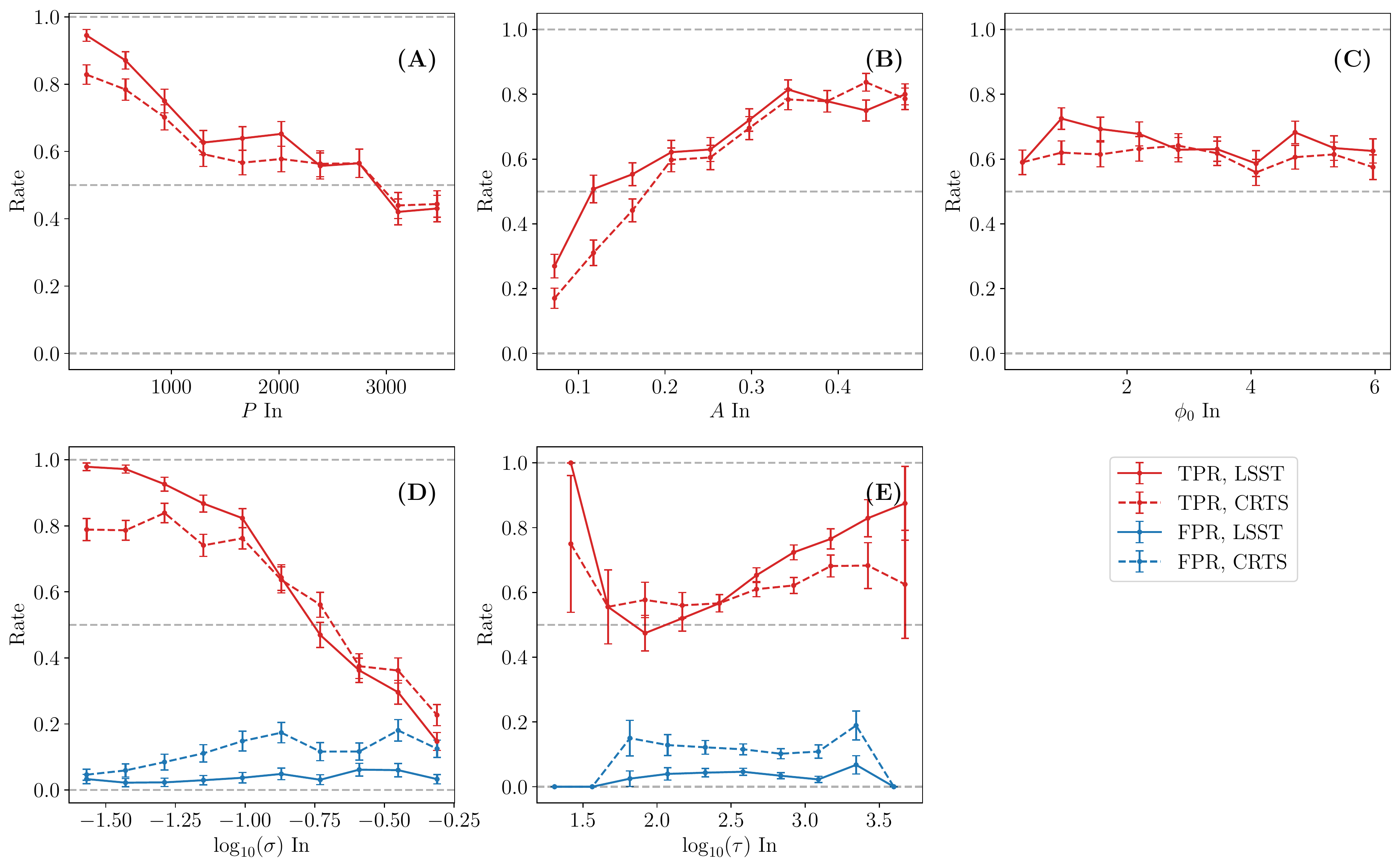}
    \caption{True-positive rates (red) and false-positive rates (blue) for LSST-like (solid lines) and CRTS-like (dashed lines) simulations, shown as a function of the input values of each parameter in the simulations. Note that false-positive rates are only shown as a function of the two DRW parameters, as there are no input sinusoids present in the false positives. The rates in each parameter bin are shown with associated uncertainties.}
    \label{fig:TPR_vs_in}
\end{figure*}

First, we simulated 1500 DRW light curves, added a randomly generated sinusoid, and then applied our model-selection scheme. 
In \autoref{fig:TPR_vs_in} we show the true-positive rate of periodic signals in the presence of DRW noise, as a function of the input parameters $P, A, \sigma, \tau$, and $\phi_0$. Here, we define the true-positive rate as the number of detected periodic signals (true positives), divided by the total number of simulated DRW+sine signals (condition positives). 
In each bin, the associated uncertainty of the rate is calculated with a binomial proportion confidence interval \citep{binomial}, where the rate can be considered as
\begin{equation}\label{eq:uncertainty}
    \frac{n_{S}}{n} \pm \frac{z}{n \sqrt{n}} \sqrt{n_{S} n_{F}}, 
\end{equation}
where $n$ is the number of trials with $n_S$ successes and $n_F$ failures, and $z$ is the $1-\alpha/2$ quantile of a normal distribution (for a 95\% confidence interval, $\alpha = 1- 0.95$).
{
We observed that our ability to detect periodicity depends both on the parameters of the sinusoid and the intrinsic DRW variability. As expected, the true-positive rate increased for high sinusoidal amplitudes and was independent of the initial phase. The true-positive rate was highest for short sinusoid periods; however, it was nonzero even for periods equal to the observation baseline, which is an unexpected improvement from \citet{vaughn}, which showed a requirement of $>2$ cycles for a sinusoid to be differentiated from a stochastic process. The true-positive rate decreased for increasing input $\sigma$; therefore, when the noise contribution became more significant, it hindered the periodicity detection, as expected. We also saw in \autoref{subsec:params} that high values of $\sigma$ (or, equivalently, low S/N) resulted in an inaccurate estimation of the parameters.
On the other hand, $\tau$ did not seem to have a significant effect on the detection rate, despite the inability to constrain large values of $\tau$, with the true-positive rate slightly increasing for longer $\tau$. Surprisingly, the overall true-positive rate varied only slightly between the two surveys.}

\begin{figure}
    \centering
    \includegraphics[width = 1\columnwidth]{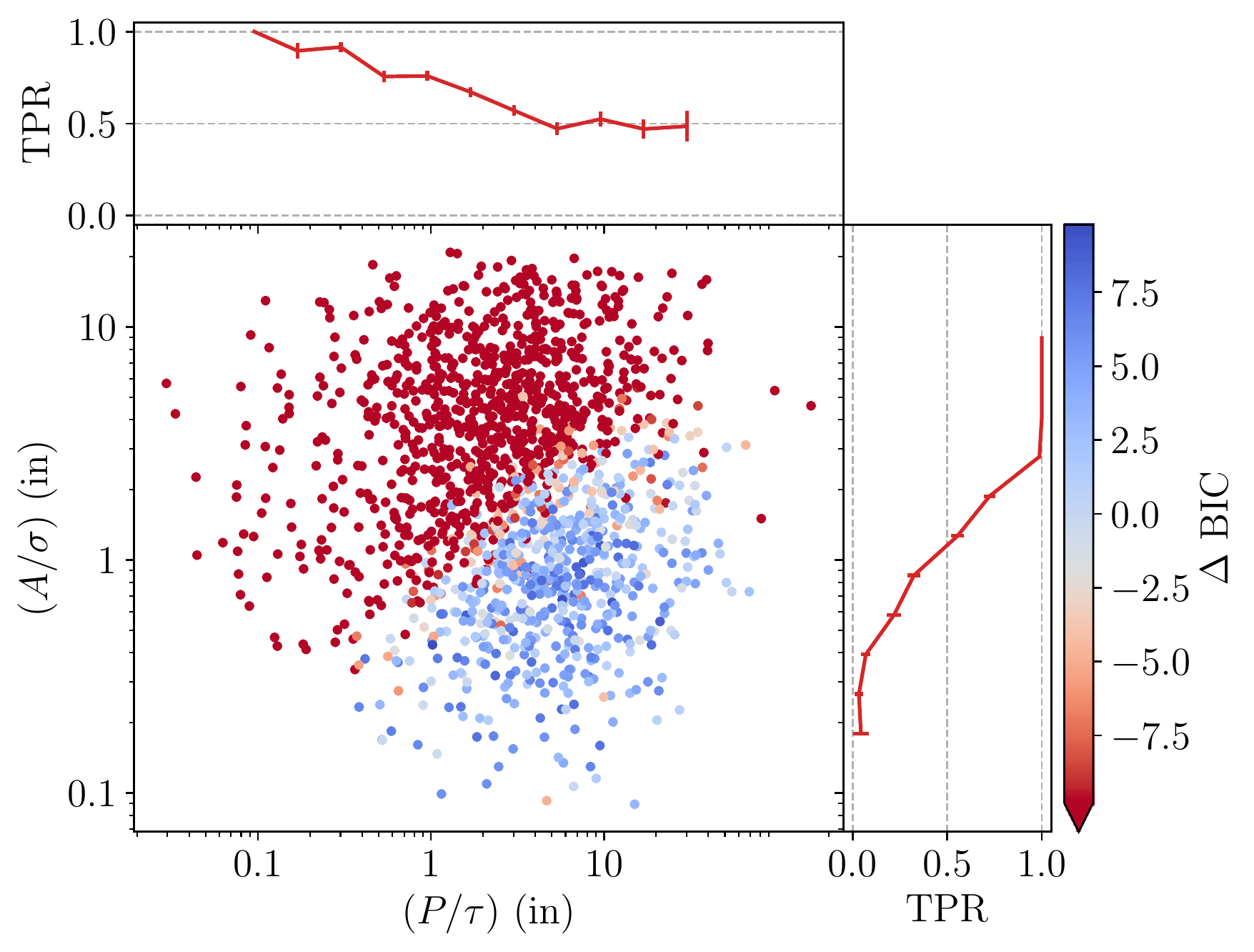}
    \caption{True-positive (red) and false-negative (blue) signals tend to lie in particular areas of parameter space in idealized LSST-like searches, when quantified by the ratios of $A/\sigma$ and $P/\tau$. The net true-positive rates integrated over $A/\sigma$ and $P/\tau$ are shown in the upper and right panels, respectively, with associated uncertainties.}
    \label{fig:p_tau_a_sig}
\end{figure}

Next, we explored how the periodicity detection rate varies as a function of the periodic parameters normalized by the noise parameters \rev{for a simulated population of quasars with a realistic distribution of $\tau$ values \citep{macleod}}. 
In \autoref{fig:p_tau_a_sig} we present the input ratios of $A/\sigma$ against $P/\tau$, colorized by the resulting $\Delta \mathrm{BIC}$. In the side panels, we track the fraction of recovered sinusoids (true-positive rate) as a function of either $A/\sigma$ (for the vertical panel) or $P/\tau$ (for the horizontal panel), again with the associated binomial uncertainty marked in each bin. 

As can be expected, the fraction of binaries recovered was highly correlated with $A/\sigma$. This value can be considered similar to an S/N; we saw that even though it was not absolutely necessary that $A>\sigma$ for a periodic signal to be detected, the detection rate dropped to $\sim$50\% when the amplitude of the sinusoid was comparable to the standard deviation of the DRW noise. The recovery fraction also clearly depends on the value of $P/\tau$, albeit less strongly than with $A/\sigma$; that is, even without considering $A/\sigma$, the periodic signal is more likely to be detected (i.e., $\Delta \mathrm{BIC}$ is lower) for smaller ratios of $P/\tau$. In terms of detectability, we see that all binary signals were identified {for small values of $P/\tau$, whereas the true-positive rate is $\sim$75\% when $P$ and $\tau$ are comparable and is further reduced to 50\% for larger values.} This is consistent with our findings in \autoref{fig:TPR_vs_in}, where we see that detectability increases for small periods and for larger values of $\tau$, although the latter correlation is weaker. The correlation of the true-positive rate with the period seen in \autoref{fig:TPR_vs_in} is fairly intuitive; a relatively weak signal can be confidently detected if the period is short and enough cycles are repeated within the data. However, it is somewhat less obvious what drives the correlation with $P/\tau$. One potential explanation is that it may be easier to detect a periodic signal if the two characteristic timescales ($P$ and $\tau$) of the light curve are fairly distinct. Otherwise, if the values are similar, they may be misidentified by the model-selection process (e.g., see \autoref{subsec:timescales}).

As a counterpoint to the previous analysis, we subsequently simulated 1500 DRW-only signals, ran our model-selection pipeline, and calculated the false-positive rate for our same detection method. This represents a scenario in which only DRW processes are occurring, and either an SMBHB is not present in the target, or it is not influencing the AGN light curve. 
Here, the false-positive rate is defined as the number of DRW-only signals identified as periodic (false positives), divided by the total number of DRW simulations (condition negatives). In \autoref{fig:TPR_vs_in}, we show the false-positive rate for both surveys as a function of input $\sigma$ and $\tau$ with blue curves, and again using the associated uncertainties calculated with \autoref{eq:uncertainty}. 
We see that the false-positive rate is significantly higher in \rev{CRTS-like simulations}, reflecting the lower measurement precision and sampling rate of the light curves, whereas in LSST the false positives are almost negligible. \rev{The overall false-positive rate for CRTS-like simulations is 12\%, while for LSST-like simulations it is 3.8\%, an improvement of \rev{approximately } an order of magnitude. The false-positive rate does not show any significant trend with $\tau$, but for CRTS-like simulations it does increase slightly for larger values of $\sigma$. This indicates that any combination of the DRW parameters is equally likely to produce a signal that can be misidentified as a sinusoid, but noisy, sparsely sampled light curves are more likely to return false positives. } We also observed that the false-positive rate did not increase for large recovered sinusoid periods, as was suggested by \citet{vaughn}; this is likely due to the use of a DRW+sine model, as opposed to a pure sinusoid.

\begin{figure}
    \centering
    \includegraphics[width = 1\columnwidth]{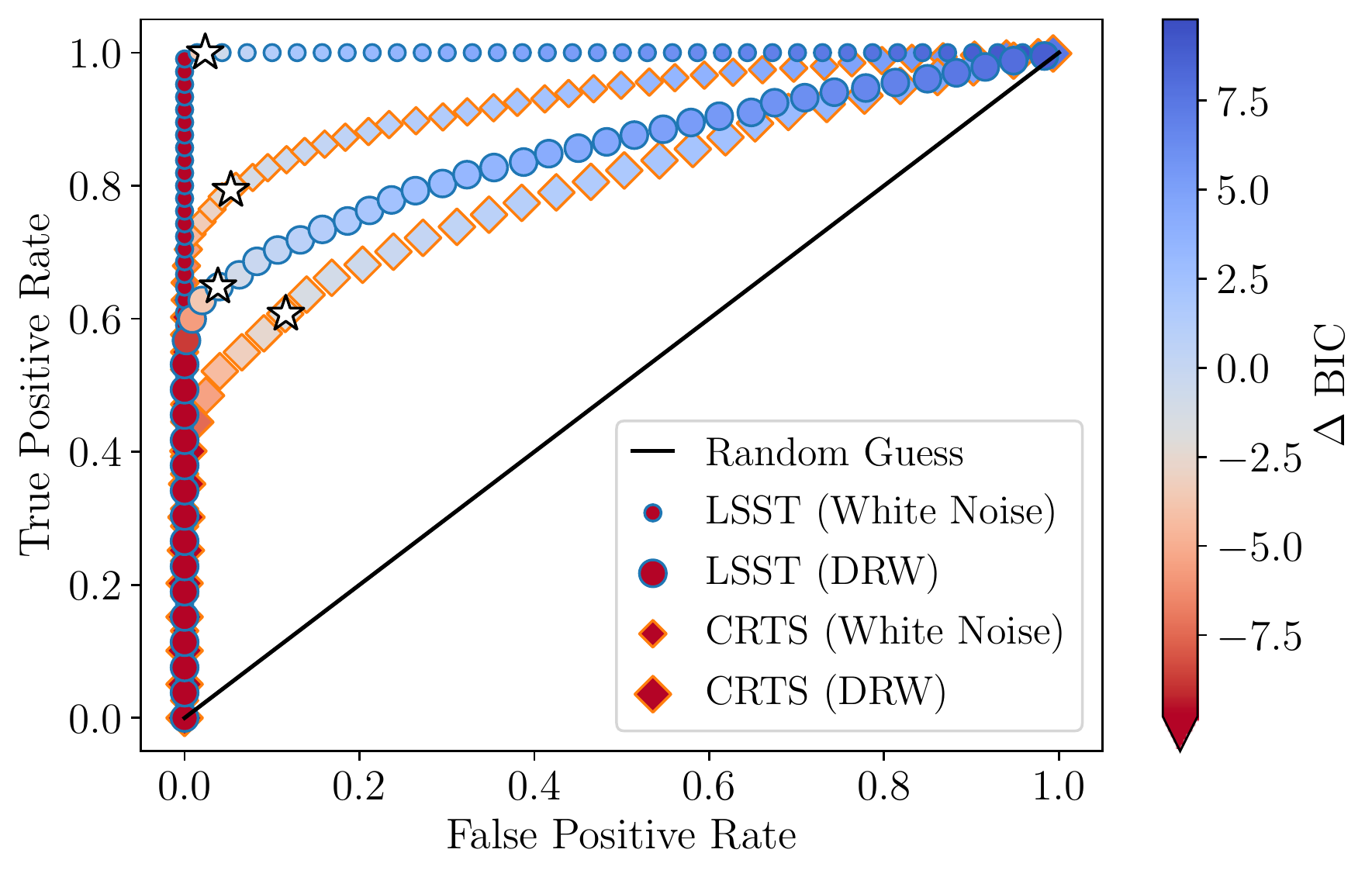}
    \caption{ROC curves for CRTS-like (orange-lined diamonds) and LSST-like (blue-lined circles) light curves. Overall, LSST can be seen to perform better than CRTS at selecting the correct model. For comparison, with our standard $\Delta\mathrm{BIC} = -2$, \rev{the true-positive rate is 64\% and the false-positive rate is 3.8\%.} Stars in the curves represent the point where $\Delta$BIC~$= -2$. Additionally, the model selection is significantly improved when white noise (small points) is present instead of a DRW process (large points), indicating that red processes are indeed a significant hindrance. }
    \label{fig:roc_color}
\end{figure}

In the above we examined the true-positive and false-positive rates as a function of the input parameters of the noise and the signal, considering a quasar to be periodic if $\Delta \mathrm{BIC}\leq-2$.  However, as is obvious from \autoref{fig:p_tau_a_sig}, these rates would be different had we chosen a different detection threshold. This is typically quantified by a receiver operating characteristic (ROC) curve, which we construct in \autoref{fig:roc_color}. More specifically, we show the true-positive rate against the false-positive rate color-coded with the threshold value for periodicity detection (i.e. the maximum $\Delta \mathrm{BIC}$ required for detection of a sinusoid within the light curve). We remind the reader that a smaller $\Delta \mathrm{BIC}$ means stronger support for the binary model. 

In \autoref{fig:roc_color}, we indicate the current threshold of $\Delta \mathrm{BIC}\leq-2$ with a star. \rev{The corresponding true-positive rate is $\sim$64\% for our LSST-like survey and $\sim$60\% our the CRTS-like survey, whereas the false-positive rate is $\sim$3.8\% for LSST-like simulations and $\sim$11\% for CRTS-like simulations. }
We note that, even though we chose this particular threshold following standard practices for model selection based on BIC, it turns out to be a reasonable threshold for both surveys. In fact, for a survey such as CRTS, it is sensible to set the threshold at a level that maximizes true positives, even if this allows some false positives. High-quality light curves are available for $\sim 10^5$ quasars, and, given that SMBHBs are relatively rare, it is manageable to pursue follow-up observations to distinguish genuine binaries from interlopers for all candidates. For LSST, on the other hand, it is critical to minimize false positives. LSST will observe millions of quasars, and follow-up of candidates needs to be significantly more selective. The colorization of \autoref{fig:roc_color} also illustrates the much larger range of $\Delta\mathrm{BIC}$ values in an LSST-like survey, as compared to CRTS. This results in a much larger number of strongly preferred signals, which will allow for a dramatically more effective ranking system for the follow-up of binary candidates.

We also show the respective ROC curves, for both CRTS-like and LSST-like simulations, for the case of periodicity on top of white noise. This allowed us to test the hypothesis that the classifier performs suboptimally due to the covariance between the sinusoid and DRW. We repeated our simulations with 1500 simulations containing white noise and a sinusoid and another 1500 with only white noise, and performed an identical model-selection procedure. Nearly all of the sinusoids were identified with accurately estimated parameters across the entire parameter space. The ROC curve for \rev{LSST-like simulations } is excellent, with close to 100\% recovery for true periodic signals and almost 0\% false detections. The ROC curve is slightly worse for CRTS due to the lower data quality. This indicates that, without the red DRW noise process included, there was no confusion, allowing the sinusoids to be identified accurately. The white-noise realization of the population, albeit unrealistic, demonstrates that the limiting factor in detecting quasar periodicity is primarily the stochastic DRW variability. 
\needrev{
\begin{table}
    \begin{tabular}{rcc}
        \hline
         & CRTS & LSST   \\
        \hline
        \hline
        DRW & 0.802 & 0.853\\
        White noise & 0.929 & 0.999 \\
        \hline
    \end{tabular}
    \caption{Area under curve (AUC) values for each of the ROC curves shown in \autoref{fig:roc}, including those with or without a DRW process.
    LSST-like surveys are expected to be a much more sensitive and reliable survey for the identification of periodicities induced by SMBHBs.}
    \label{tab:auc}
\end{table}
}

Finally, we quantitatively evaluated the performance of our method in each survey by computing the area under the ROC curve, also known as the AUC value. In general, a larger AUC value indicates a better performing classifier, as this metric equals the probability that the classifier will rank a positive simulation better than a negative one \citep[i.e. the probability that we will calculate a lower $\Delta$BIC if a sinusoid is present;][]{fawcett_roc}. 
In \autoref{tab:auc}, we summarize the AUC values for CRTS-like and LSST-like simulations both for idealized white-noise simulations and for the more realistic case that includes DRW variability. The white-noise-only ROC curve for LSST has near-perfect AUC value of 0.99, indicating that the DRW process can mask a sinusoid from the model-selection process, while white noise cannot.

\begin{figure*}
    \centering
    \includegraphics[width = 1\textwidth]{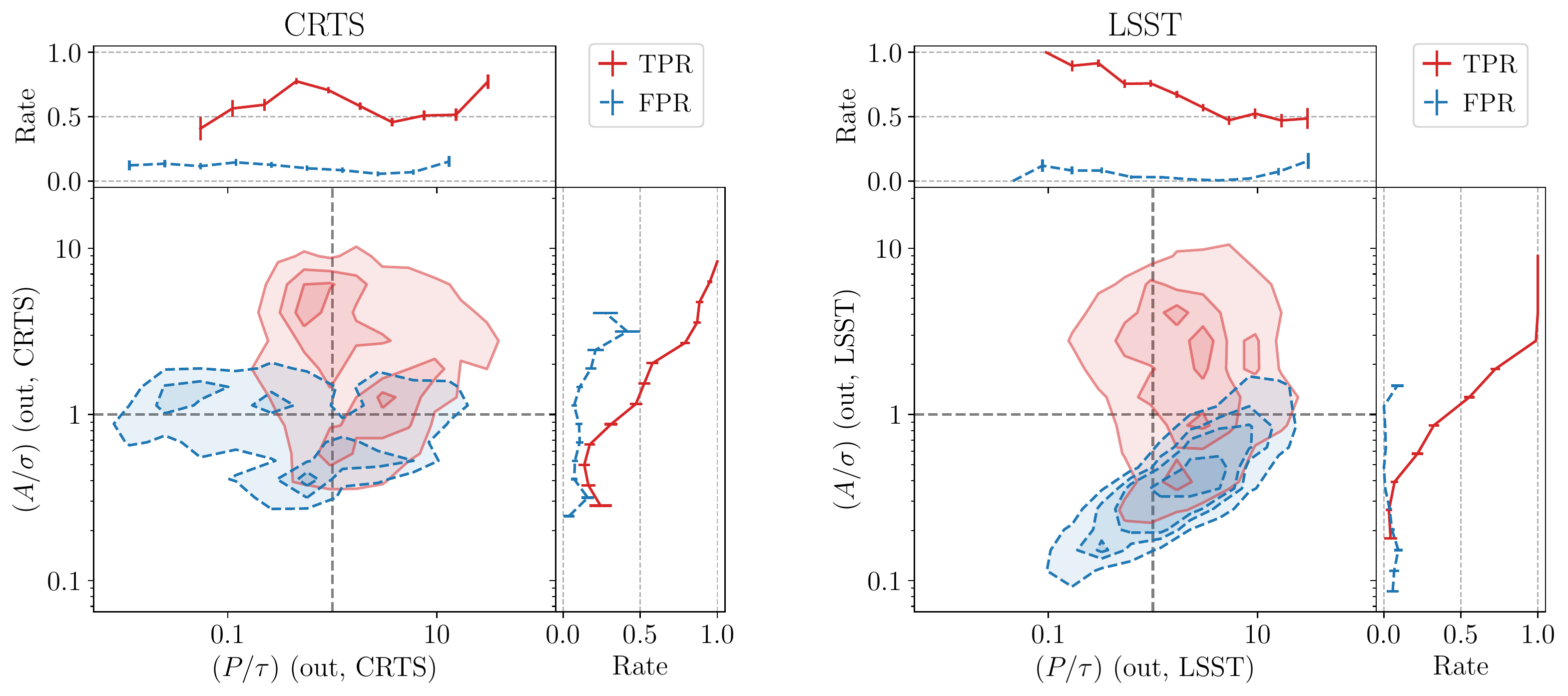}
    \caption{Simulations of DRW light curves with (red) and without (blue) a sinusoid lie in regular regions of parameter space when parametrized by the ratio of $A/\sigma$ and $P/\tau$. This makes apparent the cause of the location of false negatives in \autoref{fig:p_tau_a_sig}. It is also clear that more signals will become detectable in future surveys as cadences and baselines improve. }
    \label{fig:two_pop}
\end{figure*}

So far we have presented our results with respect to the input signals. However, in real observations we will not know the true parameters of the signals, and thus will be required to base our model-selection conclusions on the output parameters of the nested-sampling method. In \autoref{fig:two_pop}, we present the recovered parameters $A/\sigma$ versus $P/\tau$ in order to map the parts of parameter space where 
simulations with and without a sinusoid in addition to DRW noise 
are more likely to lie. 
For instance, if the DRW+sine model returns $A/\sigma>1$ \rev{for an LSST-like light curve}, it is highly likely to be a true detection regardless of $P/\tau$.
In the weak-signal regime $A/\sigma\leq1$, the two populations overlap, although, given the low number of false positives, a detected signal is more likely to be genuine periodicity. In a CRTS-like survey, it is more challenging to derive a conclusion about the validity of the detection based on the inferred parameters of the light curve, due to the higher rate of false positives. Overall, identifying periodicity in the \rev{signals with $A/\sigma>2$ } can boost our confidence that the detection is real, since no true negatives lie in this area. 

One way to quantify the 
distinction between the populations with and without a simulated sinusoid 
is with the Mahalanobis distance \citep{mahalanobis}. This metric measures the distance between a point and a distribution, measured in standard deviations of the distribution, while accounting for correlations between the data points.
For the CRTS-like observations, the median Mahalanobis distance between the two populations is \rev{$0.97$, while for the LSST-like survey, this median distance increases to $1.15$. } This indicates that, in next-generation surveys, the populations of AGN with and without sinusoidal variations will become even more clearly resolved.

\section{Discussion} \label{sec:discussion}

\subsection{Previous Work}

In this paper, we simulated CRTS-like and LSST-like light curves and used a Bayesian model selection to assess our capability to detect SMBHBs in time-domain surveys. This is the first study that explores the parameter space of sinusoidal binary signals in the presence of a DRW process, employing an array of idealized simulated data. This allowed us to examine both the detectability/completeness of binary signals and the contamination of a sample of candidates with false detections. 

We found that the sample of periodic quasars is expected to be fairly incomplete for longer-period binaries and for binaries that cause weak periodic modulations in the brightness of the AGN compared to the DRW variability. This limitation is 
caused by the stochastic variability of quasars, since in the presence of only white noise almost all the periodic signals would be detectable with nearly zero contamination. These results are independent of the time-domain survey setup. On the other hand, the false-positive rate is higher in the CRTS-like light curves compared to LSST-like ones. This suggests that the contamination of the samples of SMBHB candidates depends on the quality of the data. The reduced false-positive rate in LSST is extremely encouraging for future searches for candidate signatures of SMBHBs. This is particularly important, since LSST will observe at least 20 million quasars, and a high false-positive rate would render follow-up studies of SMBHB candidates nearly impossible.

We emphasize that even though our results provide an excellent qualitative picture of limitations and detectability trends as a function of the signal and noise parameters, they cannot be directly applied to determine the number of false positives in existing samples of SMBHB candidates \citep{Graham2015,Charisi2016,2019ApJ...884...36L}. These candidates were chosen with a different methodology, and likely suffer from distinct biases that cannot be captured by our analysis. We have already observed that, with our algorithm,
changing the detection threshold would change the true- and false-positive rate. This demonstrates that it 
will be enlightening for future systematic searches for quasar periodicity to use simulated light curves to carefully construct an ROC curve, as in our study, 
to highlight the effectiveness of the selection criteria of the search, given the specific 
survey properties.

We also note that, to date, a Bayesian model-selection method has not been applied in an extensive search for binaries. This is unsurprising, as this method is computationally demanding, and thus for a large sample of quasars (of order $10^5$ for CRTS and $10^7$ for LSST) it is practically impossible. Our idealized data sets require a few hours of CPU time per light curve to complete the model-selection analysis, and realistic data, with a larger number of associated parameters, will expand this requirement. Therefore, this method may be applied in combination with some other classifier which will make an initial preselection, and therefore significantly reduce the size of the sample. Such a complementary method will filter out most nonperiodic quasars, and thus the main requirements for it are speed and a high true-positive rate, rather than a perfect false-positive rate.

However, several Bayesian model-selection algorithms have been used in multiple studies to validate (or invalidate) the periodicity for one of the most prominent candidates, quasar PG1302-102 \citep{Dorazio2015Nature,2015Natur.518...74G,vaughn,2018ApJ...859L..12L, 2020ApJ...900..117Z}. It is intriguing that the results of these studies are not in complete agreement, neither for the best-fit parameters nor for the preferred model. This is potentially due to choices made in these analyses; for example, \citet{vaughn} introduced an extra parameter to account for poorly estimated photometric errors, \citet{2018ApJ...859L..12L} binned the light curves in wide bins of 150 days, and \citet{Dorazio2015Nature} fixed the parameters of the DRW model. This clearly illustrates the complexities of observed data sets that may not be reflected in idealized simulations, such as the ones we present in this study.

\subsection{LSST Observing Strategy and Future Improvements}
To assess the prospects of detecting SMBHBs in LSST, we simulated light curves with semi-regular sampling (evenly sampled but also adding a Gaussian error to the time stamps). As a conservative scenario for the wide-fast-deep survey, we chose a cadence of seven days, but, in reality, observations of the same source may repeat more often. We explore two additional optimistic scenarios. First, we increase the cadence to three days, and, second, we extend the observation baseline to 15 yr while keeping the cadence at seven days. We simulate both DRW and DRW+sine light curves and repeat the Bayesian model selection. 

\begin{figure}
    \centering
    \includegraphics[width = 1\columnwidth]{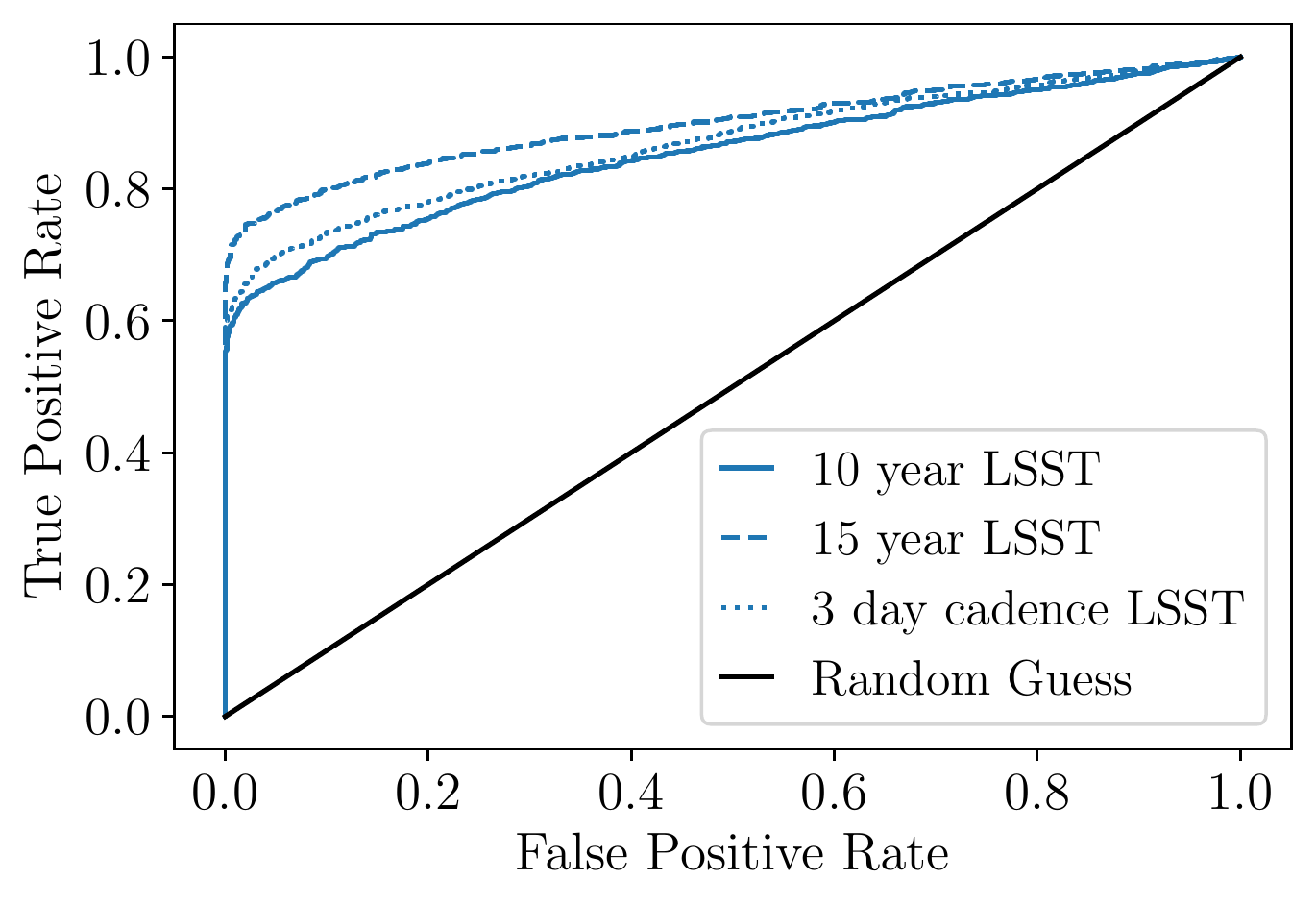}
    \caption{ROC curves for variations on our nominal LSST simulations (solid curve), including increased observation baseline (dashed curve) and increased observation cadence (dotted curve).}
    \label{fig:roc}
\end{figure}

In \autoref{fig:roc}, we present the ROC curve for these two scenarios, compared with the initial LSST-like setup as reported in \autoref{tab:surveys}. \rev{The AUC values for these two scenarios are 0.869 (three day cadence, 10 yr baseline) and 0.898 (seven day cadence, 15 yr baseline) compared to 0.854 } for our typical LSST-like simulations with a seven day cadence and 10 yr baseline. We see both from the figure and the AUC values that increasing the baseline has a positive impact in our periodicity search, allowing us to better constrain longer $\tau$ and detect longer periods at higher rates. This confirms that in preparation for LSST it is advantageous to prepare strategies that will allow us to extend the LSST light curves with already existing data by combining data from multiple surveys, such as in \citet{2018ApJ...859L..12L}. On the other hand, increased cadence does not improve our results. This is expected given that the minimum searched period is set to 30 days. We note that the higher cadence will likely significantly affect the search for short-period binaries, which are potential LISA sources \citep{xin_LISA}. In this work, we did not examine the shortest end of the period distribution, because for very short periodicities a different method may be necessary; for example, the deep coverage and more frequent sampling planned for LSST Deep Drilling Fields will certainly be beneficial for this type of source.

Moreover, even though the finalized survey strategy has not yet been decided, we recognize that our simulations are idealized for several reasons. First, 
\rev{in our LSST-like simulations, we used an average photometric uncertainty for all of the simulated light curves, even though the photometric errors are in reality magnitude dependent. In addition, we assume homoscedastic photometric errors, which is not entirely realistic; observing conditions and variability will result in heteroscedastic photometric errors, which may affect the true-/false-positive rates.} Lastly, even though the observations in the deep-wide-fast survey will repeat semi-regularly, they will rotate among six narrowband photometric filters from visit to visit. Therefore, if we consider light curves in only one photometric band, they will be significantly more sparse with $\sim$10 observations per year. The preferred route would be to combine the data in a single multiband light curve. In fact, a multiband periodogram has been developed for this purpose \citep{2015ApJ...812...18V}. However, for quasars, this process is more complicated, due to their color-dependent variability, which must be taken into account.

In future work, we intend to address several of the limitations of this current idealized study. In the near future, LSST's observing strategy (such as cadence, epoch/gap length, order of filter alternation, and frequency of observations in each photometric band, etc.) will be finalized. 
Projections of the final survey have begun to be released \citep{desc}, and as more accurate data previews, such as LSST Data Preview 0,\footnote{\url{https://rtn-001.lsst.io/}} become available over the next year, we will modify our simulations to include all the above decisions to more accurately reflect the full capabilities of the survey. To do so, we must also incorporate the magnitude dependence of the photometric errors. For this, we will simulate a more realistic quasar population, based on the quasar luminosity function, and incorporate correlations of the DRW parameters with the AGN properties (\citealt{macleod}, but see also  \citealt{2016MNRAS.459.2787K} and \citealt{graham_2017}).

In addition to improvements to the simulated observation strategies, in future work it will be critical to include a more realistic binary population and an advanced model for quasar variability \citep{2020ApJ...900..117Z, hu_tak}. More specifically, our analysis (and most searches for periodicity) assume that quasar variability is described by a DRW model. This model, albeit successful, comes with its own limitations. A future study will include advanced noise modeling and employ a continuous-time autoregressive moving-average model, which also includes quasi-periodic oscillations. For the population of SMBHBs, we randomly drew the periods and amplitudes from (log)-uniform distributions. However, binary evolution models suggest that binaries spend more time at larger separations (and longer periods) and should be more common than short-period binaries, while the amplitudes can be linked to the orbital properties of the binary (e.g., mass-ratio, and inclination for relativistic Doppler boost). We also modeled binary signals with pure sinusoids, which, while a decent approximation for a circular binary dominated by Doppler boost variability, real binaries can produce more complicated signatures. For instance, if the periodicity arises from periodic accretion or if the binary has an eccentric orbit, the light curves will significantly deviate from sinusoidal.

\subsection{Prospects for Multimessenger Observations}

Subparsec SMBHBs have remained a missing piece in the puzzle of hierarchical structure formation despite decades-long observational efforts seeking their detection. The upcoming decade is expected to bring tremendous improvements both in electromagnetic observations and in GW searches. More specifically, PTAs may be on the verge of detecting the GW background from a population of unresolvable SMBHBs \citep{12p5_gwb}. The detection of individually resolvable SMBHBs, with periods between 1~month and $\sim$10 yr \rev{(a nearly identical period range as is probed in this study) } is expected to follow soon after \citep{2015MNRAS.451.2417R,2018MNRAS.477..964K}.

On the electromagnetic side, LSST will provide a revolutionary data set for searches targeting SMBHBs. In this analysis, we have demonstrated that the unprecedented quality of the light curves will minimize the false-detection rate. 
\rev{The expected number of quasars with periodic variability that should be detectable in LSST ranges from a few hundreds to tens of thousands. More specifically, models based on the cosmological simulation \emph{Illustris} predict the detection of a few dozens candidates with Doppler boost variability and a few hundreds of candidates identified from self-lensing flares \citep{2019MNRAS.485.1579K, 2021arXiv210707522K}.  \citet{2021MNRAS.506.2408X} focused on the short-period binaries (and the potential synergy with LISA) and concluded that over 1000 binaries with period $<$200\,days should exist in the LSST database (e.g., see their Fig. 7). This population is particularly important, since it can provide insights on the expected merger rate for LISA. As we have demonstrated in this work, such binaries will be identified very reliably (see Panel A of \autoref{fig:TPR_vs_in}); the true-positive rate is the highest ($\sim$90\%) for short-period systems.}

\rev{The above developments open the possibility for combined multimessenger observations of SMBHBs \citep{2019BAAS...51c.490K}, especially since time-domain and GW experiments (like PTAs and LISA) trace overlapping populations of SMBHBs. Currently, joint observations are possible only for very high mass binaries (with masses exceeding $10^9 M_\odot$ \citep{11yrCW, charisi_prep, 3c66b}, limited by the sensitivity of PTAs; but, as PTA sensitivity improves, the common parameter space for GW and electromagnetic observations will significantly expand (e.g., see Fig. 6 in \citealt{charisi_prep}). } Incorporating priors from electromagnetic observations in the GW analysis boosts the detectability of binaries and improves parameter estimation of continuous GW searches \citep{2021arXiv210508087L}, while \citet{3c66b} showed that having a candidate to target significantly improves GW-derived upper limits on the binary chirp mass. \rev{Therefore, it is logical for GW searches to specifically target SMBHB candidates identified in time-domain surveys. This paper quantifies how LSST can produce a large number of high-quality electromagnetic SMBHB candidates, which in turn will provide a wealth of targets to search for in PTA data.}

\rev{Inversely, GW data can also provide targets for electromagnetic observations. In fact, multimessenger observations of this kind can significantly enhance the potential for discovery of long-period binaries. These systems are expected to be common, but as we demonstrated here they are also more challenging to detect based on time-domain data alone. For instance, in this paper, we have simulated sinusoids with periods of 10\,yr, equal to the baseline of LSST, and we have shown that at least of half of them would be missed in our search. A systematic search based on electromagnetic data alone would probably require multiple cycles within the LSST baseline to avoid potential confusion with the quasar noise and would exclude long-period binaries. However, we specifically decided to explore the entire parameter space, because it is possible that PTAs will detect such a binary. In that case, we can follow-up the GW detection in the LSST database and search for a binary with the identified period located within the localization volume of PTAs. We note that the PTA localization capabilities are relatively poor \citep[several hundreds or even thousands of deg$^2$,][]{taylor_16, goldstein}, and a large number of AGNs will be included in that volume. However, searching for a fixed period in combination with constraints on the total mass and the distance of the binary from the GW analysis will allow us to further filter out a significant number of candidates.}

\section{Conclusions} \label{sec:conclusions}
%
Using extensive simulations of time-domain observations of AGNs, coupled with a Bayesian model-selection and parameter-estimation framework, we have explored the capabilities of current and future surveys for SMBHB identification.
In particular, we simulated quasar light curves with DRW variability with a realistic distribution of $\sigma$ and $\tau$, as well as binary light curves with sinusoidal variability on top of a DRW process including a wide range of periods and amplitudes. We explored the likelihoods of the respective models with a Bayesian nested-sampling analysis, and determined the preferred model using the BIC. Our findings are summarized as follows:
\begin{itemize}
    \item Our ability to detect periodicity on top of DRW variability depends  on the parameters both of the sinusoid and of the noise. Short periods and high amplitudes are found at higher rates, whereas light curves with significant noise contribution (high $\sigma$) are recovered at lower rates. The input phase and $\tau$ do not appear to affect the detection rate.
    \item While our ability to discover long-period signals is decreased, about 50\% are recoverable. This is significant, because longer-period SMBHBs are expected to be more common.
    \item The true-positive rate is similar in both surveys.
    \item The incompleteness of the detectable binary signals is intrinsic due to the stochastic variability of quasars. In the presence of white noise, all periodic signals would be detectable almost independently of the data quality.
    \item The false-positive rate is higher for CRTS-like light curves and almost minimal for LSST-like data. This indicates that the high quality of LSST light curves will allow for the detection of very reliable SMBHB candidates.
    \item The false-positive rate does not depend on the input parameter of a simulated DRW signal, i.e. all DRW light curves are equally likely to produce false detections.
    \item There are parts of the parameters space where there is no significant overlap between true signals and false detections. If the recovered parameters of a light curve fall in that region (e.g., $A/\sigma>1$ for LSST-like data) it can significantly increase our confidence in the periodicity detection.
    \item If periodicity is present in a light curve, and only a DRW model is fit, the recovery of the parameters is biased.
    \item Future work will include more realistic LSST-like light curves, a wider range of binary signal models, and a physically motivated binary population.
\end{itemize}

\begin{acknowledgements}
S.B.S. and C.A.W. were supported in this work by NSF award grant Nos. 1458952 and 1815664. 
C.A.W. acknowledges support from West Virginia University through a STEM Completion Grant. 
S.B.S. is a CIFAR Azrieli Global Scholar in the Gravity and the Extreme Universe program. M.C. and S.R.T. acknowledge support from NSF grant No. AST-2007993. S.R.T also acknowledges support from an NSF CAREER Award PHY-2146016, and a Vanderbilt University College of Arts \& Science Dean's Faculty Fellowship. The authors acknowledge support from the NANOGrav Physics Frontier Center, funded by NSF Award PHY-2020265. We acknowledge the use of Thorny Flat at WVU, which is funded in part by the National Science Foundation Major Research Instrumentation Program (MRI) award No. 1726534 and WVU.
We thank Matthew Graham, who provided thoughtful comments on this manuscript.

\end{acknowledgements}

\bibliographystyle{aasjournal}
\bibliography{agn}
\end{document}